# QUANTUM COMPUTATIONS: FUNDAMENTALS AND ALGORITHMS

*S.A. Duplij[1] and I.I. Shapoval[2]*

[1] *V.N. Karazin National University, Kharkov, Ukraine,*
e-mail: sduplij@gmail.com;

[2] *National Science Center "Kharkiv Institute of Physics and Technology", Kharkiv, Ukraine;*
e-mail: ishapoval@kipt.kharkov.ua

Basic concepts of quantum information theory, principles of quantum calculations and the possibility of creation on this basis unique on calculation power and functioning principle device, named quantum computer, are concerned. The main blocks of quantum logic, schemes of quantum calculations implementation, as well as some known today effective quantum algorithms, called to realize advantages of quantum calculations upon classical, are presented here. Among them special place is taken by Shor's algorithm of number factorization and Grover's algorithm of unsorted database search.

Phenomena of decoherence, its influence on quantum computer stability and methods of quantum errors correction are described.

In the 20 century quantum physics has realized revolution in understanding the fundamental nature of the World, and in 21 - can realize revolution in the theory of computer calculations. By the 2020, taking into account the modern rates of the basic computer technologies miniaturization, we shall face that fact, that the elementary blocks of medium devices and processors of classical (Turing's) computer have reached the sizes comparable to atomic one, and cannot be correctly described within the framework of the classical theory of evaluations any more. Further development of computer technologies is impossible without change of means of classical evaluations theory based on classical physics, with a quantum apparatus, based on quantum mechanics.

Fundamental difference between characters of quantum laws and classical ones demands, generally speaking, revision of evaluations theory to realize differences in principles of quantum computer functioning, its advantages and disadvantages in comparison with a conventional computer. And even nowadays it is clear, that overcoming the miniaturization of computer devices problem and reequipping with quantum model of a data operation we obtain something much more, than the possibility of further compactification of computer's hardware components. We will get access to potentially huge computing resource, existing exclusively due to quantum mechanical properties of quantum systems (superpositions of quantum states and their entanglement) and to the quantum mechanisms, allowing us to operate with the quantum information [3].

Today it is known already several problems in solving of which quantum computer could succeed considerably in comparison with a classical computer. First of all it is a problem of large number factorization on prime factors. On conventional computers the best known algorithms of factorization are accomplished with $O\left(\exp\left[(64/9)^{1/3}(\ln N)^{1/3}(\ln \ln N)^{2/3}\right]\right)$ steps, where $N$ - input number, and $\log N$ - length of an input as the logarithm to the base, defined by scale of notation [1]. Thus, such algorithms grow exponentially with a size of input data $N$ that is an insuperable barrier to computer equipment of our day and rather long-term future even for 250-unit number. However in 1994 the algorithm for number factorization on a quantum computer was designed which is accomplished with $O((\log N)^{2+\varepsilon})$ steps, where ε is some small number [2]. It is necessary to mark, that it poses a direct threat for the majority of the modern cryptosystems (RSA, ElGamal, DiffieHellman), based on a factorization. Quantum computer will have not polynomial but nevertheless considerable advantage, above classical in a problem of searching in unsorted databases [4]. In outcome the necessary element can be found only for $O(\sqrt{N})$ calls to the database while classical searching is carried out with $O(N)$ steps that show square-law advantage of a quantum search engine.

This is the enumeration of some problems, which quantum computer promises to solve most impressively by now. Below we shall consider main principles of quantum computer operation, listed above algorithms, problems of quantum computer realization and methods of their overcoming in more detail.

A bit is the most fundamental entity of information. It is the base of conventional computer. Regardless of its physical representation, it is designed to have two distinguishable states which should have a sufficiently large energy barrier that no spontaneous transition, which would evidently be detrimental, can occur between them. It always carries two logical values as, e.g., either a "0" or a "1". It's classical. So the register, which consists of n bits, carries one of $2^n$ definite states at any given time, e.g. "$\underbrace{101...10}_{n}$".

A quantum analogue of a bit (a quantum bit, or a qubit) has, in its nature, quantum mechanical peculiarities of its behaviour. Basically any at least two-state quantum system can serve as a qubit. Its state space is the linear shell spanned on two basis vectors which are called the $|0\rangle$ and $|1\rangle$ quantum states. It is known as Hilbert one. The most essential property of a quantum state when trying to encode it is the possibility of coherence and superposition of basis states. As is known, the general state of a two-level structure quantum system is $|\Psi\rangle = \alpha|0\rangle + \beta|1\rangle$, with $|\alpha|^2 + |\beta|^2 = 1$.

Consider a register composed of $L$ qubits. It can store up to $2^L$ numbers simultaneously in a quantum superposition. Therefore, if we add more qubits to the register its capacity for storing information will increase exponentially. Then the 250-qubit register which is so small from macroscopic point of view will be capable of holding more numbers than there are atoms in the known Universe (If anything, this understates the amount of quantum information that they hold, for in general, the elements of a superposition are present in continuously variable proportions, each with its own phase angle as well.) Even so if we measure the register's content, we will see only one of those numbers. However, now we can start doing some non-trivial quantum computation, for once the register is prepared in a superposition of many different numbers, we can perform mathematical operations on all of them at once.

The contents of the $L$-qubit registers can be thought of as a $2^L$-dimensional complex vector. An algorithm for a quantum computer must initialize this vector in some specified form (dependent on the design of the quantum computer). In each step of the algorithm, that vector is modified by multiplying it by a unitary matrix. The matrix is determined by the physics of the device. The unitary character of the matrix ensures the matrix is invertible (so each step is reversible).

Upon termination of the algorithm, the $2^L$-dimensional complex vector stored in the register must be somehow read off from the qubit register by a quantum measurement. However, by the laws of quantum mechanics, that measurement will yield a random $L$ bit string (and it will destroy the stored state as well). This random string can be used in computing the value of a function because (by design) the probability distribution of the measured output bit string is skewed in favor of the correct value of the function. By repeated runs of the quantum computer and measurement of the output, the correct value can be determined, to a high probability, by majority polling of the outputs. In brief, quantum computations are probabilistic.

A quantum algorithm is implemented by an appropriate sequence of unitary operations. Note that for a given algorithm, the operations will always be done in exactly the same order. There is no "IF THEN" statement to vary the order, since there is no way to read the state of a qubit before the final measurement. There are, however, conditional gate operations such as the controlled NOT gate, or CNOT [5, 10].

A quantum algorithm is any physical process which utilizes characteristically quantum effects to perform useful computational tasks. It is convenient to formalize the description of these quantum computational processes n terms of a model which closely parallels the formalism of classical computation. In essence, the memory bits of the computer are qubits either than bits and the elementary operations are unitary transformations, each operating on a fixed finite number of qubits, rather than the Boolean operations of classical computation. It may be argued [6] that a model of this type suffices to describe any general quantum physical process. Any computer is required to operate by 'finite means' i.e. it is equipped only with the possibility of applying any operation of some finite fixed set of basic unitary operations. Any other unitary operation that we may need in an algorithm must be built (or rather approximated to sufficient accuracy) out of these basic building blocks by concatenating their action on selected qubits. It may be shown [7] that various quite small collections of unitary operations (so-called 'universal sets' of operations) suffice to approximate any unitary operation on any number of qubits to arbitrary accuracy.

One of the most useful and significant consequences of this formalism is that it provides a way of assessing the complexity of a computational task (again by paralleling concepts from classical computational complexity theory).

In the study of quantum algorithms it is of paramount interest to find polynomial-time algorithms for problems where no classical polynomial time algorithm is known, i.e. we wish to demonstrate that quantum effects may give rise to an exponential speedup in running time over classical information processing. We will describe the situation in which this occurs on the Shor's algorithm. We will also describe the quantum searching algorithm which provides a square root speedup over any classical algorithm, rather than an exponential speedup.

Shor's algorithm is a quantum algorithm for factoring a number $N$ in $O((\log N)^3)$ time and $O(\log N)$ space, named after Peter Shor [2].

The algorithm is significant because it implies that RSA, a popular public-key cryptography method, might be easily broken, given a sufficiently large quantum computer. Shor's algorithm can crack RSA in polynomial time [8].

Like many quantum computer algorithms, Shor's algorithm is probabilistic: it gives the correct answer with high probability, and the probability of failure can be decreased by repeating the algorithm. However, since a proposed answer (in particular primality) is polynomial time verifiable, the algorithm can be modified to work in expected polynomial time with zero error.

Shor's algorithm was discovered in 1994 by Peter Shor, but the classical part was known before, it is credited to G.L. Miller. Seven years later, in 2001, it was demonstrated by a group at IBM, which factored 15 into 3 and 5, using a quantum computer with 7 qubits.

The problem we are trying to solve is that, given an integer $N$, we try to find another integer $p$ between 1 and $N$ that divides $N$.

Shor's algorithm consists of two parts:

1) A reduction of the factoring problem to the problem of order-finding, which can be done on a classical computer.

2) A quantum algorithm to solve the order-finding problem.

The classical part is as follows:

1) Pick a random number $a < N$

2) Compute $\gcd(a, N)$ (gcd – greatest common divisor). This may be done using the Euclidean algorithm.

3) If $\gcd(a, N) \neq 1$, then there is a nontrivial factor of $N$, so we are done.

4) Otherwise, use the period-finding subroutine (below) to find $r$, the period of the following function:
$$f(x) = a^x \mod(N),$$
i.e. the smallest integer $r$ for which $f(x+r) = f(x)$.

5) If $r$ is odd, go back to step 1.

6) If $a^{r/2} \equiv -1 \pmod{N}$, go back to step 1.

7) The factors of $N$ are $\gcd(a^{r/2} \pm 1, N)$. We are done.

Now consider the quantum part: period–finding subroutine:

1) Start with a pair of input and output qubit registers with $\log_2 N$ qubits each, and initialize them to $N^{-1/2} \sum_x |x\rangle |0\rangle$, where $x$ runs from 0 to $N-1$.

2) Construct $f(x)$ as a quantum function and apply it to the above state, to obtain
$$U_{QFT}|x\rangle = N^{-1/2} \sum_y e^{-2\pi i x y / N} |y\rangle.$$

This leaves us in the following state:
$$N^{-1} \sum_x \sum_y e^{-2\pi i x y / N} |y\rangle |f(x)\rangle.$$

3) Perform a measurement. We obtain some outcome $y$ in the input register and $f(x_0)$ in the output register. Since $f$ is periodic, the probability of measuring some pair $y$ and $f(x_0)$ is given by
$$\left| N^{-1} \sum_{x: f(x)=f(x_0)} e^{-2\pi i x y / N} \right|^2 = N^{-2} \left| \sum_b e^{-2\pi i (x_0 + rb) y / N} \right|^2.$$

Analysis now shows that this probability is higher, the closer $yr/N$ is to an integer.

4) Turn $y/N$ into an irreducible fraction, and extract the denominator $r'$, which is a candidate for $r$.

5) Check if $f(x) = f(x + r')$. If so, we are done.

6) Otherwise, obtain more candidates for $r$ by using values near $y$, or multiples of $r'$. If any candidate works, we are done.

7) Otherwise, go back to step 1 of the subroutine.

The algorithm is composed of two parts. The first part of the algorithm turns the factoring problem into the problem of finding the period of a function, and may be implemented classically. The second part finds the period using the inverse quantum Fourier transform, and is responsible for the quantum speedup.

So in the first stage the factors from period are obtained. The integers less than $N$ and coprime with $N$ form a finite group under multiplication modulo $N$, which is typically denoted $(Z/NZ)^x$. By the end of step 3, we have an integer $a$ in this group. Since the group is finite, $a$ must have a finite order $r$, the smallest positive integer such that $a^r \equiv 1 \mod N$

Therefore, $N \mid (a^r - 1)$. Suppose we are able to obtain $r$, and it is even. Then
$$a^r - 1 = (a^{r/2} - 1)(a^{r/2} + 1) \equiv 0 \mod N$$
$$\Rightarrow N \mid (a^{r/2} - 1)(a^{r/2} + 1).$$

$r$ is the smallest positive integer such that $a^r \equiv 1$, so $N$ cannot divide $(a^{r/2} - 1)$. If $N$ also does not divide $(a^{r/2} + 1)$, then $N$ must have a nontrivial common factor with each of $(a^{r/2} - 1)$ and $(a^{r/2} + 1)$. This supplies us with a factorization of $N$. If $N$ is the product of two primes, this is the only possible factorization.

The second part is devoted to finding the period. Shor's period-finding algorithm relies heavily on the ability of a quantum computer to be in many states simultaneously. To compute the period of a function $f$, we evaluate the function at all points simultaneously.

Quantum physics does not allow us to access all this information directly, though. A measurement will yield only one of all possible values, destroying all others. Therefore we have to carefully transform the superposition to another state that will return the correct answer with high probability. This is achieved by the inverse quantum Fourier transform [9].

Shor thus had to solve three "implementation" problems. All of them had to be implemented "fast", which means that they can be implemented with a number of quantum gates that is polynomial in $\log N$.

1. Create a superposition of states. This can be done by applying Hadamard gates [10] to all qubits in the input register. Another approach would be to use the quantum Fourier transform (see below).

2. Implement the function $f$ as a quantum transform. To achieve this, Shor used repeated squaring for his modular exponentiation transformation. It is important to note that this step is more difficult to implement than the quantum Fourier transform, in that it requires ancillary qubits and substantially more gates to accomplish.

3. Perform an inverse quantum Fourier transform. By using controlled rotation gates and Hadamard gates Shor designed a circuit for the quantum Fourier transform that uses just $O\left((\log N)^2\right)$ gates.

After all these transformations a measurement will yield an approximation to the period $r$. For simplicity assume that there is a $y$ such that $yr/N$ is an integer. Then the probability to measure $y$ is 1. To see that we notice that then $e^{-2\pi i b y r / N} = 1$ for all integers $b$. Therefore the sum whose square gives us the probability to measure $y$ will be $N/r$ since $b$ takes roughly $N/r$ values and thus the probability is $1/r^2$. There are $ry$ such that $yr/N$ is an integer and also $r$ possibilities for $f(x_0)$, so the probabilities sum to 1 [P.W. Shor, Polynomial-Time Algorithms for Prime Factorization and Discrete Logarithms on a Quantum Computer, SIAM J. Sci. Statist. Comput. 26 (1997) 1484].

Now we will review in brief the Grover's algorithm [4]. It is a quantum algorithm for searching an unsorted database with $N$ entries in $O(\sqrt{N})$ time and using $O(\log N)$ storage space (see big O notation). It was invented by Lov Grover in 1996.

Classically, searching an unsorted database requires a linear search, which is $O(N)$ in time. Grover's algorithm, which takes $O(\sqrt{N})$ time, is the fastest possible quantum algorithm for searching an unsorted database. It provides "only" a quadratic speedup, unlike other quantum algorithms, which may provide exponential speedup over their classical counterparts. However, even quadratic speedup is considerable when $N$ is large.

Like many quantum computer algorithms, Grover's algorithm is probabilistic in the sense that it gives the correct answer with high probability. The probability of failure can be decreased by repeating the algorithm. (An example of a deterministic quantum algorithm is the Deutsch-Jozsa algorithm, which always produces the correct answer with probability one.)

Below, we present the basic form of Grover's algorithm, which searches for a single matching entry.

Consider an unsorted database with $N$ entries. The algorithm requires an $N$-dimensional state space $H$, which can be supplied by $\log_2 N$ qubits.

Let us number the database entries by 0, 1, ... ($N$-1). Choose an observable, $\Omega$, acting on $H$, with $N$ distinct eigenvalues whose values are all known. Each of the eigenstates of $\Omega$ encode one of the entries in the database, in a manner that we will describe. Denote the eigenstates (using bra-ket notation) as $\{|0\rangle, |1\rangle, ..., |N-1\rangle\}$ and the corresponding eigenvalues by $\{\lambda_0, \lambda_1, ..., \lambda_{N-1}\}$.

We are provided with a unitary operator, $U_\omega$, which acts as a subroutine that compares database entries according to some search criterion. The algorithm does not specify how this subroutine works, but it must be a quantum subroutine that works with superpositions of states. Furthermore, it must act specially on one of the eigenstates, $|\omega\rangle$, which corresponds to the database entry matching the search criterion. To be precise, we require $U_\omega$ to have the following effects: $U_\omega |\omega\rangle = -|\omega\rangle$ and $U_\omega |x\rangle = |x\rangle$ for all $x \neq \omega$. Our goal is to identify this eigenstate $|\omega\rangle$, or equivalently the eigenvalue $\omega$, that $U_\omega$ acts specially upon.

The steps of Grover's algorithm are as follows [4]:

1. Initialize the system to the state $|s\rangle = \frac{1}{\sqrt{N}} \sum_x |x\rangle$.

2. Perform the following "Grover iteration" $r(N)$ times. The function $r(N)$ is described below.
   a. Apply the operator $U_\omega$;
   b. Apply the operator $U_s = 2|s\rangle\langle s| - I$.

3. Perform the measurement $\Omega$. The measurement result will be $\lambda_\omega$ with probability approaching 1 for $N>>1$. From $\lambda_\omega$, $\omega$ may be obtained.

Our initial state is $|s\rangle = \frac{1}{\sqrt{N}} \sum_x |x\rangle$. Consider the plane spanned by $|s\rangle$ and $|\omega\rangle$. Let $|\omega^x\rangle$ be a ket in this plane perpendicular to $|\omega\rangle$. Since $|\omega\rangle$ is one of the basis vectors, the overlap is $\langle \omega | s \rangle = \frac{1}{\sqrt{N}}$. In geometric terms, there is an angle $(\pi/2 - \theta)$ between $|\omega\rangle$ and $|s\rangle$, where $\theta$ is given by $\cos\left(\frac{\pi}{2} - \theta\right) = \frac{1}{\sqrt{N}}$ and $\sin\theta = \frac{1}{\sqrt{N}}$.

The operator $U_\omega$ is a reflection at the hyperplane orthogonal to $|\omega\rangle$; for vectors in the plane spanned by $|s\rangle$ and $|\omega\rangle$, it acts as a reflection at the line through $|\omega^x\rangle$. The operator $U_s$ is a reflection at the line through $|s\rangle$. Therefore, the state vector remains in the plane spanned by $|s\rangle$ and $|\omega\rangle$ after each application of $U_s$ and after each application of $U_\omega$, and it is straightforward to check that the operator $U_s U_\omega$ of each Grover iteration step rotates the state vector by an angle of $2\theta$ toward $|\omega\rangle$.

We need to stop when the state vector passes close to $|\omega\rangle$; after this, subsequent iterations rotate the state vector away from $|\omega\rangle$, reducing the probability of obtaining the correct answer. The number of times to iterate is given by $r$. In order to align the state vector exactly with $|\omega\rangle$, we need:

$$\frac{\pi}{2} - \theta = 2\theta r, \quad r = \frac{1}{4}\left(\frac{\pi}{\theta} - 2\right).$$

However, $r$ must be an integer, so generally we can only set $r$ to be the integer closest to $\frac{1}{4}\left(\frac{\pi}{\theta} - 2\right)$. The angle between $|\omega\rangle$ and the final state vector is $O(\theta)$, so the probability of obtaining the wrong answer is $O(1 - \cos^2\theta) = O(\sin^2\theta)$. For $N>>1$, $\theta \approx N^{-1/2}$, so $r \to \frac{\pi\sqrt{N}}{4}$.

Furthermore, the probability of obtaining the wrong answer becomes $O(1/N)$, which goes to zero for large $N$.

There are a number of practical difficulties in building a quantum computer, and thus far quantum computers have only solved trivial problems. To summarize the problem from the perspective of an engineer, one needs to solve the challenge of building a system which is isolated from everything except the measurement and manipulation mechanism. Furthermore, one needs to be able to turn off the coupling of the qubits to the measurement so as to not decohere the qubits while performing operations on them.

One major problem is keeping the components of the computer in a coherent state, as the slightest interaction with the external world would cause the system to decohere. This effect causes the unitary character (and more specifically, the invertibility) of quantum computational steps to be violated. Decoherence times for candidate systems, in particular the transverse relaxation time $T_2$ (terminology used in NMR and MRI technology, also called the dephasing time), typically range between nanoseconds and seconds at low temperature [11]. The issue for optical approaches are more difficult as these timescales are orders of magnitude lower and an often cited approach to overcome it uses optical pulse shaping approach. Error rates are typically proportional to the ratio of operating time to decoherence time, hence any operation must be completed much quicker than the decoherence time. If the error rate is small enough, it is possible to use quantum error correction, which corrects errors due to decoherence, thereby allowing the total calculation time to be longer than the decoherence time. An often cited (but rather arbitrary) figure for required error rate in each gate is $10^{-4}$. This implies that each gate must be able to perform its task 10,000 times faster than the decoherence time of the system.

Meeting this scalability condition is possible for a wide range of systems. However the use of error correction brings with it the cost of a greatly increased number of required qubits. The number required to factor integers using Shor's algorithm is still polynomial, and thought to be between $L^4$ and $L^6$, where $L$ is the number of bits in the number to be factored. For a 1000 bit number, this implies a need for $10^{12}$ to $10^{18}$ qubits. Fabrication and control of this large number of qubits is non-trivial for any of the proposed designs.

One approach to the stability-decoherence problem is to create a topological quantum computer [12] with anyons, quasi-particles used as threads and relying on knot theory to form stable logic gates.

Quantum error correction [13] is for use in quantum computing to protect quantum information from errors due to decoherence and other quantum noise. Quantum error correction is essential for fault-tolerant quantum computation which is designed to deal not just with noise on stored quantum information, but also with faulty quantum gates, faulty quantum preparation, and faulty measurements.

Classical error correction employs redundancy: The simplest way is to store the information multiple times, and — if these copies are later found to disagree — just take a majority vote; e.g. If a bit has been copied three times but now one bit says 0 but two bits say 1, then it is probable that the original state was three 1s, and a single error occurred, than that originally it was three 0s and two errors occurred, though that could have happened. Although copying is not possible with quantum information, due to the no-cloning theorem [14], the information of one qubit may be spread onto several (physical) qubits by using a quantum error correcting code. Such encoded quantum information is protected, as in classical error correcting codes, against errors of a limited form.

As in classical error correcting codes, a syndrome measurement can determine whether a qubit has been corrupted, and if so, which one. What is more, the outcome of this operation (the syndrome) tells us not only which physical qubit was affected, but also, in which of several possible ways it was affected. The latter is counter-intuitive at first sight: Since noise is arbitrary, how can the effect of noise be one of only few distinct possibilities? In most codes, the effect is either a bit flip, or a sign (of the phase) flip, or both (corresponding to the Pauli matrices $X$, $Z$, and $Y$). The reason is that the measurement of the syndrome has the projective effect of a quantum measurement. So even if the error due to the noise was arbitrary, it can be expressed as a superposition of basis operations—the error basis (which is here given by the Pauli matrices and the identity). The syndrome measurement "forces" the qubit to "decide" for a certain specific "Pauli error" to "have happened", and the syndrome tells us which, so that we can let the same Pauli operator act again on the corrupted qubit to revert the effect of the error.

The crucial point is that the syndrome measurement tells us as much as possible about the error that has happened, but nothing at all about the value that is stored in the logical qubit — as otherwise the measurement would destroy any quantum superposition of this logical qubit with other qubits in the quantum computer.

We have discussed some aspects and capabilities of quantum computations. It is clear that it will be incredibly difficult to realize quantum computer technically. We could do that if we were sure that the expected benefit will surpass our efforts. Listed above applications are one of the most promising up-to-date algorithms which can be realized on quantum computers, but not only one. But it seems that the most effective use of quantum machines will be in quantum systems simulation which will find application in chemistry, materials science, nanotechnology, biology and medicine. So, we have to search new possible applications and we have understanding of wide interdisciplinary efforts necessity to realize quantum computers as the fastest computational devices in the world.

## REFERENCES

1. A.M. Odiyzko. *The future of integer factorization:* Preprint At&T Bell Laboratories. 1995, 16 p.
2. P. Shor. Polynomial-time algorithms for prime factorization and discrete logs on a quantum computer //*SIAM J. Sci. Statist. Comput.* 1997, v. 26, 1484 p.
3. R. Feynman. Simulating physics with computers //*Int. J. Theoret. Phys.* 1982, v. 21, p. 467-488.
4. L.K. Grover. A fast quantum mechanical algorithm for database search //*Proc. 28$^{th}$ Annual ACM*

*Symposium on the Theory of Computing.* 1996, p. 212-219.
5. C. Monroe et. al. Demonstration of a Fundamental Quantum Logic Gate //*Physical Review Letters*. 1995, v. 75, p. 4714-4717.
6. D. Deutsch. Quantum theory, the Church-Turing principle, and the universal quantum computer //*Proc. R. Soc. London A*., 1985, v. 400, p. 97-117.
7. D. Deutsch, A. Barenco, A. Ekert. *Universality in quantum computation* //*Proc. Roy. Soc. London A* 1995, v. 449, p. 669-677.
8. R. Rivest et al. *On digital signatures and public-key cryptosystems:* Preprint MIT/LCS/TR-212, MIT Laboratory for Computer Science, 1979.
9. M.A. Nielsen, I.L. Chuang. *Quantum Computation and Quantum Information.* Cambridge: "Cambridge University Press", 2000, 665 p.
10. A. Barenco et al. Elementary gates for quantum computation. //*Phys.Rev. A*. 1995, v. 52, p. 3457-3488.
11. D.P. DiVincenzo. Quantum computation //*Science*. 1995, v. 270, N 5234, p. 255-261.
12. G.P. Collins. Computing with Quantum Knots //*Scientific American*. April 2006, p. 57-63.
13. P.W. Shor. Scheme for reducing decoherence in quantum computer memory //*Phys. Rev. A*. 1995, v. 52, p. 2493–2496.
14. W.K. Wootters, W.H. Zurek. A Single Quantum Cannot be Cloned //*Nature*. 1982, v. 299, p. 802-803.

## КВАНТОВЫЕ ВЫЧИСЛЕНИЯ: ОСНОВЫ И АЛГОРИТМЫ

*С.А. Дуплий, И.И. Шаповал*

Рассмотрены основные концепции квантовой теории информации, принципы квантовых вычислений и возможность создания на их основе уникального по вычислительной мощности и принципу функционирования устройства – квантового компьютера. Представлены основные блоки квантовой логики, схемы реализации квантовых вычислений, а также известные сегодня эффективные квантовые алгоритмы, которые призваны воплотить преимущества квантовых вычислений над классическими. Среди них особое место занимают алгоритм Шора – факторизации чисел и алгоритм Гровера – поиска в неупорядоченных базах данных. Описано явление декогеренции, её влияние на стабильность квантового компьютера и методы коррекции квантовых ошибок.

## КВАНТОВІ ОБЧИСЛЕННЯ: ОСНОВИ ТА АЛГОРИТМИ

*С.А. Дуплій, І.І. Шаповал*

Розглянуто основні концепції квантової теорії інформації, принципи квантових обчислень та можливість створення на їх основі унікального по обчислювальній потужності та принципу функціювання пристрою – квантового комп'ютера. Представлені основні блоки квантової логіки, схеми впровадження квантових обчислень, а також відомі сьогодні ефективні квантові алгоритми, що покликані втілити переваги квантових обчислень над класичними. Серед них особливе місце займають алгоритм Шора – факторизації чисел та алгоритм Гровера – пошуку в невпорядкованих базах даних. Описано явище декогеренції, її вплив на стабільність квантового комп'ютера та методи корекції квантових помилок.